\documentclass[runningheads]{llncs}
\usepackage[T1]{fontenc}
\usepackage{graphicx}
\pdfoutput=1

\usepackage{placeins}
\usepackage{setspace}
\usepackage{cite}
\usepackage{amsmath}
\usepackage{mathtools}
\usepackage{siunitx}
\usepackage{subcaption}
\usepackage{bm}
\usepackage{xspace}
\usepackage{hyperref}
\usepackage{cleveref}
\usepackage[toc, page]{appendix}
\ifpdf
\hypersetup{
  colorlinks=true,
  linkcolor=blue,
  citecolor=blue,
  urlcolor=blue,
  pdftitle={Wide-scale Monitoring of Satellite Lifetimes: \linebreak
Pitfalls and a Benchmark Dataset},
  pdfauthor={}
}
\fi

\usepackage{xargs}                      \usepackage[pdftex,dvipsnames]{xcolor}  \usepackage[colorinlistoftodos,prependcaption,textsize=scriptsize]{todonotes}
\newcommandx{\td}[2][1=]{\todo[linecolor=cyan,backgroundcolor=cyan!5,bordercolor=cyan,#1]{#2}}
\newcommandx{\ds}[2][1=]{\todo[linecolor=teal,backgroundcolor=teal!5,bordercolor=teal,#1]{DS: #2}}
\newcommandx{\jm}[2][1=]{\todo[linecolor=violet,backgroundcolor=violet!5,bordercolor=violet,#1]{JM: #2}}
\newcommandx{\mr}[2][1=]{\todo[linecolor=OliveGreen,backgroundcolor=OliveGreen!5,bordercolor=OliveGreen,#1]{MR: #2}}

\newcommand{\figRef}[1]{\autoref{#1}}
\newcommand{\eqRef}[1]{(\autoref{#1})}
\newcommand{\secRef}[1]{\autoref{#1}}

\newcommand{\eg}{{\em e.g., }}
\newcommand{\ie}{{\em i.e., }}

\begin{document}
\title{Optimal Proposal Particle Filters for Detecting Anomalies and Manoeuvres from
Two Line Element Data}
\titlerunning{Particle Filters for Manoeuvres from Two Line Element Data}
\author{David P. Shorten\inst{1} \and
John Maclean\inst{1} \and
Melissa Humphries\inst{1} \and
Yang Yang\inst{2} \and
Matthew Roughan\inst{1}}

\authorrunning{D. Shorten et al.}
\institute{The University of Adelaide, Adelaide SA 5000, Australia \and
The University of New South Wales, Sydney NSW 2052, Australia}
\maketitle              
\begin{abstract}
    Detecting anomalous behaviour of satellites is an important goal
    within the broader task of space situational awareness. The Two
    Line Element (TLE) data published by NORAD is the only
    widely-available, comprehensive source of data for satellite
    orbits. We present here a filtering approach for detecting
    anomalies in satellite orbits from TLE data. Optimal proposal
    particle filters are deployed to track the state of the
    satellites' orbits. New TLEs that are unlikely given our belief of
    the current orbital state are designated as anomalies. The change
    in the orbits over time is modelled using the SGP4 model with
    some adaptations. A model uncertainty is derived to handle the errors in
    SGP4 around singularities in the orbital elements.
    The proposed techniques are
    evaluated on a set of 15 satellites for which ground truth is
    available and the particle filters are shown to be superior at
    detecting the subtle in-track and cross-track manoeuvres in the
    simulated dataset, as well as providing a measure of uncertainty
    of detections.
\end{abstract}
 
\vspace{-3mm}
\section{Introduction}

Space Situational Awareness (SSA) is an important task for both governments and commercial operators of
satellites~\cite{kennewell2013overview}.
Knowledge of the orbits of satellites allows for mission planning, resource management and strategic countermeasures.
This task is of ever-increasing importance given the growing number of space objects and the growing technological
importance of satellites~\cite{zhao2022overview}.
This work focuses on a sub-goal of SSA, which is the detection of anomalies in satellite orbits. These anomalies
can correspond to satellite collisions, malfunctions or manoeuvres.

Detailed tracking measurements, such as optical and radar data, are
not publicly available for most satellites, but the Two-Line Element (TLE)~\cite{celesTrak} data provided by the
North American Aerospace Defense Command (NORAD) provides long-term
details of the orbits of a large number
of satellites.

Anomalies in the TLE catalogue can be detected using data-centric time-series analysis, and this approach has
been demonstrated to be effective
~\cite{kelecy2007satellite, roberts2021geosynchronous, wang2021gaussian}.
However, TLE data is intended to be used in conjunction with SGP4~\cite{hoots1980spacetrack}, a non-linear orbital
model which accounts for non-elliptical components. The standard use-case for SGP4 is to derive a precise
satellite position at
given time points from the most recent TLE set~\cite{vallado2006revisiting}. However, the orbital parameters published
in TLEs are the mean Keplerian elements.
That is, they are Keplerian elements that represent an averaging of the orbit of the satellite which excludes short
term periodic variations~\cite{vallado2001fundamentals}.
SGP4 also propagates these mean elements before incorporating non-Keplerian components.
It has recently been argued that the most
appropriate use of SGP4 in the context of analysing the TLE catalogue is to use it to propagate the mean orbital
elements and ignore precise satellite positions and the added non-Keplerian components of the
orbit~\cite{shorten2022wide, shorten2023wide}.

We present a method for detecting orbital anomalies in TLE data.
The orbit of each satellite is tracked by filtering TLE observations  in the NORAD  catalogue. Observations which are
improbable, given our belief of the tracked mean orbit are then designated as anomalous.
A particle filtering approach is used  to handle the non-linearity of the SGP4 model. Adaptations are made
to the standard implementation of this model to make it compatible with filtering. Further, a study is made of the
errors in this model caused by singularities in the orbital elements. A suitable model uncertainty is derived to
ameliorate the affect of these errors.

To the best of the authors' knowledge, this is the first work that presents an approach to satellite orbit anomaly
detection from TLE data based on tracking the underlying orbital state, operating on the mean orbital elements with
no non-Keplerian components incorporated. This is an important step given the increasing importance of our contested
space environment and the fact that TLE data is the only widely-available data source for space objects.

The proposed techniques are evaluated on a set of 15 satellites for
which independent ground truth is
available~\cite{shorten2022wide}. The filtering approach outperformed
existing approaches~\cite{li2018new,
  li2019maneuver,decoto2015technique,
  mukundan2021simplified,zhao2014method}, in particular on more subtle
manoeuvres and those such as cross-track manoeuvres that are less
commonly seen and have not been thoroughly considered in the past.

\vspace{-3mm}
\section{Filtering Techniques}
\label{sec:filtering_techniques}

Although satellite orbits are often described as elliptical, they
follow a more complex, non-linear path. Orbital models have been
developed over many years: we use the best widely-used model, the SGP4
model~\cite{hoots1980spacetrack}. The challenge here is to perform
non-linear filtering on this complex model to find unusual events.

The elliptical approximation of a satellites' orbit is described by 6
orbital parameters, called the {\em mean elements}.  These elements
are denoted  at the
$k$\textsuperscript{th} epoch by the vector $\mathbf{x}_k = \{e'_k,
i'_k, n'_k,\Omega'_k, \omega'_k, M'_k \}^T$, where $e$ is eccentricity,
$i$ inclination, $n$ mean motion, $\Omega$ the longitude of the
ascending node, $\omega$ the argument of perigee and $M$ the mean
anomaly. In a naive model all of these except the mean anomaly would
be constant. The primes emphasize that they are mean elements and thus
subject to more complex dynamics. 

Observations are the mean elements recorded in the TLE data, denoted
by $\mathbf{y}_k = \{e'_{\text{TLE}, k}, i'_{\text{TLE}, k},
n'_{\text{TLE}, k},\Omega'_{\text{TLE}, k}, \omega'_{\text{ TLE}, k},
M'_{\text{TLE}, k} \}^T$, and we denote a {\em trajectory} of
observations over epochs $1$ through $k$ by by the sequence
$\mathbf{y}_{1:k}$.
 
We seek to estimate the distribution of the orbital elements, given
the observations, \ie $p\left( \mathbf{x}_{k} \mid \mathbf{y}_{1:k}
\right)$. Knowledge of this distribution and the orbit-evolution model
allows estimation of the distribution of future observations from the
existing trajectory, \ie $p\left( \mathbf{y}_{k+1} \mid
\mathbf{y}_{1:k} \right)$.  An observation is designated as anomalous
if it is unlikely under this distribution.

Using the SGP4 model, the satellite's orbit obeys
\begin{equation}
    \label{eq:dynamics}
    \mathbf{x}_k = f_{\text{SGP4}} \left( \mathbf{x}_{k-1}, t_{k-1}, t_{k} \right) + \boldsymbol{\xi},
\end{equation}
where $\boldsymbol{\xi} \sim \mathcal{N}\left(0, \mathbf{Q}\right)$ is
Gaussian model error, $f_{\text{SGP4}}(\cdot)$ incorporates the physics of
orbital mechanics, and $t_k$ is the epoch associated with
$\mathbf{x}_k$.
Note that we are specifically referring to the component of SGP4 describing the evolution of the
mean orbital elements, without adding non-Keplerian effects (the osculating elements). As was
recently demonstrated in~\cite{shorten2023wide} it is advantageous to focus on
these mean elements for anomaly detection. We are also
using a non-standard implementation of SGP4 in that the mean motion is not converted from the Kozai to Brouwer
formulations~\cite{vallado2006revisiting}. Rather, this conversion is done as a pre-processing step across the TLE set.

We assume Gaussian uncertainty in TLE observations and so
\begin{equation}
    \label{eq:observation}
    \mathbf{y}_k = \mathbf{x}_{k} + \boldsymbol{\eta_k},
\end{equation}
where $\boldsymbol{\eta_k} \sim \mathcal{N}\left(0, \mathbf{R}\right)$
is the measurement noise, which has covariance $\mathbf{R}$.

In many respect the problem, when composed in this manner resembles a
standard tracking/filtering application, but we have found it is very
important to (i) get this exact formulation correct (\eg Kozai vs
Brouwer formulations matter), and (ii) to estimate reasonable
covariances, noting that orbital measurement errors are not truly
Gaussian. These factors mean that the type of filter does seem to
matter. We test two approaches.

\vspace{-3mm}
\subsection{Bootstrap Particle Filter (BS-PF)}
\label{sec:filtering_techniques:bootstrap}

The following description of the filters draws on~\cite{arulampalam2002tutorial} and~\cite{speekenbrink2016tutorial}.
Tracking draws on a set of particles with parameters and associated
weights $\{\mathbf{x}_{k}^i, w_k^i\}_{i=1}^{N}$. Each particle
represents a sample of the orbital state at epoch $k$. A proposal
distribution $q$ is used to generate samples, and 
 weights are given by 
$
w_k^i
\propto
p\left(\mathbf{x}_{0:k}^i \mid \mathbf{y}_{1:k} \right)
/ \\
q\left(\mathbf{x}_{0:k}^i\mid \mathbf{y}_{1:k} \right)
\label{eq:weights}, 
$
where $p$ comes from the model, and $q$ the proposal distribution. 
The trajectory  of the
$i$\textsuperscript{th} particle is given by $\mathbf{x}_{0:k}^i$ up to the $k$\textsuperscript{th}
epoch. We can then approximate the distribution of orbital
trajectories as
$    p \left( \mathbf{x}_{0:k} \mid \mathbf{y}_{1:k} \right)
    \approx
    \Sigma_{i=1}^{N} w_k^i \delta \left( \mathbf{x}_{0:k} - \mathbf{x}_{0:k}^i \right) .
    \label{eq:posterior_density}
$
It is useful to write $w_k^i$ in terms of the weight at the previous epoch:
$
w_k^i
\propto
w_{k-1}^i
p\left(\mathbf{y}_{k} \mid \mathbf{x}_{k}^i \right)
p\left(\mathbf{x}_{k}^i \mid \mathbf{x}_{k-1}^i \right)
/
q\left(\mathbf{x}_{k}^i \mid \mathbf{x}_{0:k-1}^i, \mathbf{y}_{1:k} \right).
\label{eq:weights_update}
$
As we are primarily interested in tracking the current orbital state $\mathbf{x}_{k}$, as opposed to the orbital
trajectory $\mathbf{x}_{0:k}$, we make changes to focus on the current state. Using the fact that the model dynamics
in \eqref{eq:dynamics} are Markovian, and the measurement errors $ \boldsymbol{\eta_k}$ in \eqref{eq:observation} are
independent, we factorise the distribution of orbit trajectories to obtain the
probability density of the
current state as
\begin{equation}
    p \left( \mathbf{x}_{k} \mid \mathbf{y}_{1:k} \right)
    \approx
    \Sigma_{i=1}^{N} w_k^i \delta \left( \mathbf{x}_{k} - \mathbf{x}_{k}^i \right) .
    \label{eq:posterior_density_current}
\end{equation}
The set of particles is initialised by adding Gaussian noise to the first TLE in the set. That is, it is constructed by
drawing $N$ samples from $\mathcal{N}\left(\mathbf{y}_1, \mathbf{S}\right)$. The initial weights are all set to $1/N$.

We then proceed iteratively through the epochs of the TLE set by drawing a new sample from the proposal distribution
$q\left(\mathbf{x}_{k}^i \mid \mathbf{x}_{0:k-1}^i, \mathbf{y}_{1:k} \right)$
for each particle $i$ at the next epoch $k$.

The bootstrap  particle filter uses the proposal distribution based
on the model, \ie
$q_{\text{boot}}\left(\mathbf{x}_{k}^i \mid \mathbf{x}_{0:k-1}^i, \mathbf{y}_{1:k} \right) =
p\left(\mathbf{x}_{k}^i \mid \mathbf{x}_{k-1}^i \right)$.
We  therefore draw the $i$\textsuperscript{th} particle at the $k$\textsuperscript{th} epoch from
$
\mathcal{N}\left(
        \mathbf{x}_{k}^i \, ; \,
        f_{\text{SGP4}} \left(
            \mathbf{x}_{k-1}^i, t_{k-1}, t_{k}
        \right),
        \mathbf{Q}
    \right).
$
The weights are updated by
$
    w_k^i = w_{k-1}^i p\left(\mathbf{y}_{k} \mid \mathbf{x}_{k}^i \right),
$
before being normalised to sum to 1.

Weight degeneracy, where the weight of one particle approaches 1 and
all others 0, is a problem here. We avoid degeneracy through
regularization. The level of degeneracy is monitored by tracking the
effective sample size, which can be estimated by $
\hat{N}_{\text{effective}, k} = 1 / \Sigma_i \left(w_k^i\right)^2.  $
When $\hat{N}_{\text{effective}, k} / N < \tau_r$, regularisation is
triggered. The particles are first resampled by sampling from the
current particle set in proportion to their weights.  This resampling
is performed according to the {\em systematic} strategy
~\cite{speekenbrink2016tutorial}.  A single number $u_k$ is drawn from
the uniform distribution between 0 and 1. Then, for each $j$ from $1$
through $N$, we find the particle $i$ such that $\Sigma_{m = 1}^{i-1}
w_k^m \leq (j-1)/N + u_k < \Sigma_{m = 1}^{i} w_k^m$ and this
$i$\textsuperscript{th} particle is added to the resampled particle
set.  In cases of more severe degeneracy, many of the resampled
particles will be copies of a small number of original particles. To
mitigate this, we add noise so that the resampled particles are
effectively drawn from a smooth approximation of $p \left(
\mathbf{x}_{k} \mid \mathbf{y}_{1:k} \right)$. Specifically, we use
the approximation $ p \left( \mathbf{x}_{k} \mid \mathbf{y}_{1:k}
\right) \approx \Sigma_{i=1}^{N} w_k^i K_h \left( \mathbf{x}_{k} -
\mathbf{x}_{k}^i \right) , $ where $K_h$ is the Gaussian kernel with
bandwidth $h = N^{\left( -1 / (n_x + 4)
  \right)}$~\cite{musso2001improving} (where $n_x= 6$ throughout this
paper). In practice, this is achieved by, for each particle $i$,
drawing a random sample $\epsilon^i$ from the standard normal
distribution and perturbing the $i$\textsuperscript{th} particle by
$h\mathbf{D_k}\epsilon^i$, where $\mathbf{D_k}$ is a root of the
empirical covariance matrix of the weighted ensemble of
particles~\cite{musso2001improving}.

\vspace{-3mm}
\subsection{Optimal Proposal Particle Filter (OP-PF)}
\label{sec:filtering_techniques:optimal_proposal}

The weight degeneracy problem is exacerbated if the proposal distribution $q$ is not a close match to the
distribution $p$. In the context of tracking mean orbital elements, we know that this will be the case for the
bootstrap filter, where we choose our proposal distribution to be
$q_{\text{boot}}\left(\mathbf{x}_{k}^i \mid \mathbf{x}_{0:k-1}^i, \mathbf{y}_{1:k} \right) =
p\left(\mathbf{x}_{k}^i \mid \mathbf{x}_{k-1}^i \right) =
\mathcal{N}\left(\mathbf{x}_{k}^i \, ; \, f_{\text{SGP4}} \left( \mathbf{x}_{k-1}^i, t_{k-1}, t_{k} \right),
\mathbf{Q}\right)$.

Given the substantial errors in the SGP4 model for propagating mean elements, as explored in \secRef{sec:uncertainty}, we do not expect
that this proposal will closely match the target distribution $p\left( \mathbf{x}_{k} \mid \mathbf{y}_{1:k} \right)$.

A superior proposal distribution is 
$q_{\text{opt}}\left(\mathbf{x}_{k} \mid \mathbf{x}_{0:k-1}, \mathbf{y}_{1:k} \right) =
p\left( \mathbf{x}_{k} \mid \mathbf{x}_{k-1}, \mathbf{y}_{k} \right)$.
This proposal is known to be optimal in the sense that it minimises the variance of the
resulting particle weights from a large class of proposal distributions~\cite{arulampalam2002tutorial,leeuwen2019particle}.

This optimal proposal can be evaluated analytically in the case of nonlinear dynamics and linear observations with
Gaussian uncertainty (see \cite{doucet2000sequential,snyder2011particle}). Specifically, in our case where the observation matrix is the identity,
~\cite{arulampalam2002tutorial} shows:
\begin{equation}
    q_{\text{opt}}\left(\mathbf{x}_{k}^i \mid \mathbf{x}_{0:k-1}^i, \mathbf{y}_{1:k}^i \right) =
\mathcal{N} \left(\mathbf{x}_{k}^i \, ; \, \mathbf{m}_k^i, \mathbf{\Sigma}_{p} \right),
\end{equation}
where
$
    \mathbf{\Sigma}_{p}^{-1} = \mathbf{Q}^{-1} + \mathbf{R}^{-1}
$
and
$
    \mathbf{m}_{k}^i = \mathbf{\Sigma}_{p} \left(\mathbf{Q}^{-1} f_{\text{SGP4}} \left( \mathbf{x}_{k-1}^i, t_{k-1}, t_{k} \right) +
    \mathbf{R}^{-1} \mathbf{y}_{k}  \right) \;.
$

Apart from this change to the proposal density, we run the optimal proposal filter in the same manner as specified
for the bootstrap filter in \secRef{sec:filtering_techniques:bootstrap}.

\section{Anomaly detection}
\label{sec:anomaly_detection}

We set our anomaly statistic $s_{\text{anom}, k}$ as the negative predictive
density of the subsequent observation:
$
    s_{\text{anom}, k} \coloneqq - p\left( \mathbf{y}_{k+1} \mid \mathbf{y}_{1:k}\right).
$
Observations at the $k$\textsuperscript{th} epoch are designated as anomalous if $s_{\text{anom}, k}$ is greater than
a chosen threshold $\tau_{\text{anom}}$.
The predictive density of the subsequent orbital state can be evaluated as:
$
    p\left( \mathbf{x}_{k+1} \mid \mathbf{y}_{1:k}\right) =
    \int p\left(\mathbf{x}_{k+1} \mid \mathbf{x}_{k}\right) p\left( \mathbf{x}_{k} \mid \mathbf{y}_{1:k} \right) d\mathbf{x}_{k}.
$
Making use of \eqRef{eq:posterior_density_current}, this can be approximated by:
$
    p\left( \mathbf{x}_{k+1} \mid \mathbf{y}_{1:k}\right) \approx
    \Sigma_{i=1}^{N} w_k^i  p\left(\mathbf{x}_{k+1} \mid \mathbf{x}_{k}^i\right).
$
As our observation and model uncertainty are normal and independent, the predictive density of subsequent
observations can be approximated as:
\begin{equation}
    \label{eq:predictive_density}
    p\left( \mathbf{y}_{k+1} \mid \mathbf{y}_{1:k}\right) \approx
    \Sigma_{i=1}^{N} w_k^i
    \mathcal{N}\left(\mathbf{y}_{k+1} \, ; \, f_{\text{SGP4}} \left( \mathbf{x}_{k}^i, t_{k}, t_{k+1} \right),
    \mathbf{Q} + \mathbf{R}\right)
\end{equation}

As direct computation of \eqRef{eq:predictive_density} regularly leads to numerical underflow,
we instead compute the negative logarithm of the predictive density by first computing the logarithm of
each term in the sum and then applying the log-sum-exp trick~\cite{blanchard2021accurately}.

Large manoeuvres can lead to subsequent TLE observations being situated sufficiently far from the existing particle
ensemble that the filters diverge. To avoid this, when the negative logarithm of the predictive density in
\eqRef{eq:predictive_density} rises above a threshold $\tau_{\text{shift}}$, then the particle ensemble is shifted so
that the new mean of the ensemble is equal to the observation $\mathbf{y}_k$.

\section{Model and Observation Uncertainty}
\label{sec:uncertainty}

\begin{figure}
    \centering
    \begin{subfigure}[t]{0.49\linewidth}
        \centering
        \includegraphics[width = 1.0\linewidth]{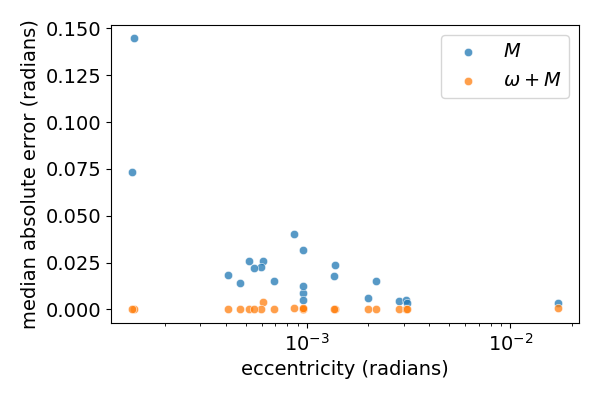}
        \caption{Non-equatorial satellites\\ ($i > \SI{0.01}{\radian}$ )}
        \label{fig:sgp4_errors:vs_eccentricity}
    \end{subfigure}
    \begin{subfigure}[t]{0.49\linewidth}
        \centering
        \includegraphics[width = 1.0\linewidth]{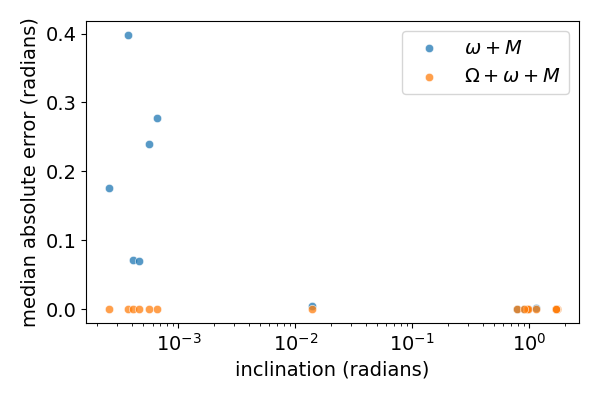}
        \caption{All satellites (equatorial and non-equatorial).}
        \label{fig:sgp4_errors:vs_inclination}
    \end{subfigure}
    \caption{Plots showing the median absolute error of propagations for combinations of certain elements, against
    the inclination or eccentricity.}
    \label{fig:sgp4_errors}
\end{figure}

Before filtering, we must determine the covariance matrices, $\mathbf{Q}$ and $\mathbf{R}$,
associated with the model and observation uncertainties (see Equations \ref{eq:dynamics} and \ref{eq:observation}).

This uncertainty was determined by analysing the residuals of SGP4 predictions.
We found substantial errors in the prediction of  $\omega$, $M$ and $\Omega$.
The substantial errors in $\omega$ and $M$ have already been highlighted in the
literature~\cite{san2017hybrid}, where it was noted that the prediction of the sum of these elements
was much more accurate than the prediction of the individual elements.
This type of error will not impact the main intended purpose of
SGP4 (predicting precise satellite positions in Cartesian coordinates~\cite{hoots1980spacetrack}) due to
singularities in the Keplerian orbital elements at small eccentricity~\cite{vallado2001fundamentals}. In this regime,
the ellipse become almost circular and the orientation of the ellipse given by $\omega$ has less impact on the final
satellite position.

We confirmed the findings of~\cite{san2017hybrid} by downloading the TLE sets of 60 randomly chosen satellites
from the list of satellites provided by the Union of Concerned Scientists (UCS)~\cite{ucs}.
Each TLE pair in each set was propagated to the subsequent epoch and the error in the propagation was then
found. \figRef{fig:sgp4_errors:vs_eccentricity} shows large median absolute errors in the propagation of $M$ and far
smaller errors in the propagation of $\omega + M$.

The Keplerian orbital elements also contain singularities when the inclination is 0 or 90 degrees. We focus on the
small inclination case as this is far more common.
\figRef{fig:sgp4_errors:vs_inclination} shows large median absolute errors in the propagation of $\omega + M$ at
small inclinations (below 0.01 radians) and smaller errors for  $\Omega + \omega + M$.

NORAD does not publish information on the uncertainty associated with each TLE record~\cite{vallado2012two}.
Further, there is minimal published work on the accuracy of SGP4 when used for propagating mean orbital elements.
We estimate $Q$ and $R$ using an approach similar to what has been used for
estimating the uncertainty associated with the combination of TLE data and the full SGP4 model for providing precise
satellite positions~\cite{geul2017tle}. We find the residuals
$\mathbf{r}_{k} = \mathbf{x}_{k} - f_{\text{SGP4}}\left( \mathbf{x}_{k-1}^i, t_{k-1}, t_{k} \right)$
associated with propagating each TLE to the subsequent epoch across the entire TLE set for a given satellite.
We then find the maximum likelihood estimate of the
covariance of these residuals (assuming a mean of 0), $\mathbf{\hat{\Sigma}}_{\text{res}}$. This will be a combined
estimate of the uncertainty of both our model (SGP4) and our observations. As both of these uncertainties are
assumed to be additive and independent (see Equations \ref{eq:dynamics} and \ref{eq:observation}), $\mathbf{\hat{\Sigma}}_{\text{res}}$
is an estimate of the sum  $\mathbf{Q} + \mathbf{R}$. There is no sensible way to separate out this uncertainty.
We therefore act conservatively and set both $\mathbf{Q}$ and $\mathbf{R}$ as
$\mathbf{\hat{\Sigma}}_{\text{res}}$.

This initial setting of $\mathbf{Q}$ and $\mathbf{R}$ needs to be modified to incorporate our knowledge of the errors
associated with SGP4 discussed above.
These model errors are non-random, in the sense that similar orbital elements result in a similar direction and
magnitude of model error, leading to the divergence of the particle filters. This divergence can be prevented by
inflating the model uncertainty for the elements with large errors.

Inflating the model uncertainty produces an overly wide proposal distribution
$q\left(\mathbf{x}_{k} \mid \mathbf{x}_{0:k-1}, \mathbf{y}_{1:k} \right)$, resulting in weight degeneracy. We
partially address this by tailoring the structure of the model uncertainty covariance $\mathbf{Q}$ to better match
our knowledge of the structure of errors in the SGP4 model. The model covariance is modified such that it
represents a distribution where $\xi_{\omega} + \xi_{M} = 0$ in the case of non-equatorial orbits
and $\xi_{\Omega} + \xi_{\omega} + \xi_{M} = 0$ in the case of equatorial orbits
($\boldsymbol{\xi} \sim \mathcal{N}\left(0, \mathbf{Q}\right)$).

The resulting covariance matrix for the non-equatorial case can be written
\begin{equation}
    \mathbf{Q}_{\text{non-eq.}} =
    \text{diag} \left( \hat{\boldsymbol{\sigma}}_{\text{non-eq.}}^2 \right)^{\left(\frac{1}{2}\right)}
    \begin{pmatrix}
        1                 & \hat{\rho}_{e, i} & \hat{\rho}_{e, n} & \hat{\rho}_{e, \Omega} & \hat{\rho}_{e, \omega} & \hat{\rho}_{e,M}\\
        \hat{\rho}_{i, e} & 1                 & \hat{\rho}_{i, n} & \hat{\rho}_{i, \Omega} & \hat{\rho}_{i, \omega} & \hat{\rho}_{i,M}\\
        \hat{\rho}_{n, e} & \hat{\rho}_{n, i} & 1                 & \hat{\rho}_{n, \Omega} & \hat{\rho}_{n, \omega} & \hat{\rho}_{n,M}\\
        \hat{\rho}_{\Omega, e} & \hat{\rho}_{\Omega, i} & \hat{\rho}_{\Omega, n} & 1 & \hat{\rho}_{\Omega, \omega} & \hat{\rho}_{\Omega,M}\\
        \hat{\rho}_{\omega, e} & \hat{\rho}_{\omega, i} & \hat{\rho}_{\omega, n} & \hat{\rho}_{\omega, \Omega} & 1 & -1\\
        \hat{\rho}_{M, e} & \hat{\rho}_{M, i} & \hat{\rho}_{M, n} & \hat{\rho}_{M, \Omega} & -1 & 1\\
    \end{pmatrix}
    \text{diag} \left( \hat{\boldsymbol{\sigma}}_{\text{non-eq.}}^2 \right)^{\left(\frac{1}{2}\right)} ,
    \label{eq:Q_non-eq}
\end{equation}
where $\hat{\boldsymbol{\sigma}}_{\text{non-eq.}}^2$ is the diagonal of the maximum likelihood covariance
matrix of the residuals resulting from propagating the mean elements in each satellite's TLE set to the
subsequent epoch
 ($\mathbf{\hat{\Sigma}}_{\text{res}}$), with an inflation factor $\alpha$ applied to $\omega$ and $M$. Specifically:
$
    \hat{\boldsymbol{\sigma}}_{\text{non-eq.}}^2 =
    \{
    \hat{\sigma}^2_e,
    \hat{\sigma}^2_i,
    \hat{\sigma}^2_n,
    \hat{\sigma}^2_{\Omega},
    \alpha \hat{\sigma}^2_{\omega},
    \alpha \hat{\sigma}^2_M
    \}^T.
$
Each $\hat{\rho}_{x, y}$ is computed from the estimated covariance matrix $\mathbf{\hat{\Sigma}}_{\text{res}}$.
Similarly, the covariance matrix for the equatorial case can be written as
\begin{equation}
    \mathbf{Q}_{\text{eq.}} =
    \text{diag} \left( \hat{\boldsymbol{\sigma}}_{\text{eq.}}^2 \right)^{\left(\frac{1}{2}\right)}
    \begin{pmatrix}
        \ddots & & & \vdots \\
        & 1 & -\frac{1}{2} & -\frac{1}{2}\\
        & -\frac{1}{2} & 1 & -\frac{1}{2}\\
        \dots & -\frac{1}{2} & -\frac{1}{2} & 1\\
    \end{pmatrix}
    \text{diag} \left( \hat{\boldsymbol{\sigma}}_{\text{eq.}}^2 \right)^{\left(\frac{1}{2}\right)},
\end{equation}
where
$
    \hat{\boldsymbol{\sigma}}_{\text{eq.}}^2 =
    \{
    \hat{\sigma}^2_e,
    \hat{\sigma}^2_i,
    \hat{\sigma}^2_n,
    \alpha \hat{\sigma}^2_{\Omega},
    \alpha \hat{\sigma}^2_{\omega},
    \alpha \hat{\sigma}^2_M
    \}^T.
$ We have only shown the bottom right block of the correlation matrix, as the rest of the matrix is identical to the
one in \eqRef{eq:Q_non-eq}.

Estimated from the SGP4 propagation residuals, $\mathbf{\hat{\Sigma}}_{\text{res}}$ already includes a strong
negative correlation between $\omega$ and $M$ in the non-equatorial case and between $\Omega$, $\omega$ and $M$ in the
equatorial case. This correlation needs to be removed for the observation uncertainty covariance matrix
$\mathbf{R}$.
To remove this negative correlation, and also for implementation convenience, we model the observations as
being independent in each element. That is
$
    \mathbf{R} =
    \text{diag} \left( \hat{\boldsymbol{\sigma}}_{\text{res}}^2 \right)
$
where $\hat{\boldsymbol{\sigma}}_{\text{res}}^2$ is the diagonal of $\mathbf{\hat{\Sigma}}_{\text{res}}$, with no
inflation factor applied. 
\vspace{-3mm}
\section{Experiments}
\label{sec:experiments}

Each method being evaluated assigns an anomaly statistic $s_{\text{anom}, k}$ to each epoch $k$.
In practical application, some threshold $\tau_{\text{anom}}$ would be chosen such that epochs where
$s_{\text{anom}, k} > \tau_{\text{anom}}$ would be classified as anomalous. We evaluate each method at all possible
thresholds to obtain precision-recall curves. This evaluation is performed using an
event-matching scheme \cite{zhao2021event}. Each predicted anomaly is matched to
the closest manoeuvre within time $t_{\text{matching}}$. Manoeuvres with a matching prediction are counted as true
positives and those without one are considered false negatives. Similarly, predictions which are
not matched are considered false positives.

Both particle filters used the following parameters: a number of particles of $N = 500$, a variance inflation factor
of $\alpha =3$, a threshold on $N_{\text{effective}}$ for regularisation of $\tau_r = 0.2$ and a threshold on the
negative log predictive density for ensemble shift of $\tau_{\text{shift}} = \SI{1e1}{}$. All code for running these
experiments will be made available on publication.

We construct a baseline which fits within the class of methods for TLE manoeuvre
detection which propagate the TLE to the subsequent epoch(s) using SGP4, before comparing
with the measured TLEs in these subsequent epochs~\cite{li2018new, mukundan2021simplified, zhao2014method}.
We implement possibly the most straightforward
instance of this class, which simply propagates each TLE to the subsequent epoch, before making a comparison. The
anomaly statistic associated with epoch $k$ is then
\begin{equation}
    \label{eq:baseline_anomaly_statistic}
    s_{\text{anom, baseline}, k} =
    \lVert
    f_{\text{SGP4}} \left( \mathbf{x}_{k-1}, t_{k-1}, t_{k} \right) -
    \mathbf{x}_{k}
    \rVert.
\end{equation}

\vspace{-3mm}
\subsection{Evaluation on Benchmark Dataset}
\label{sec:experiments:benchmark}

\begin{figure}[ht!]
    \centering
    \begin{subfigure}[t]{0.65\linewidth}
        \centering
        \includegraphics[width = 1.0\linewidth]{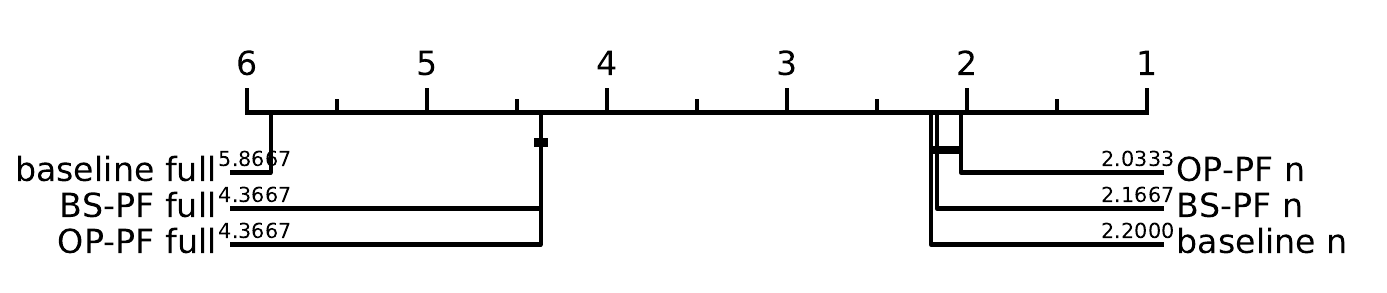}
        \caption{Benchmark dataset}
        \label{fig:cd_plots:benchmark}
    \end{subfigure}
    \begin{subfigure}[t]{0.65\linewidth}
        \centering
        \includegraphics[width = 1.0\linewidth]{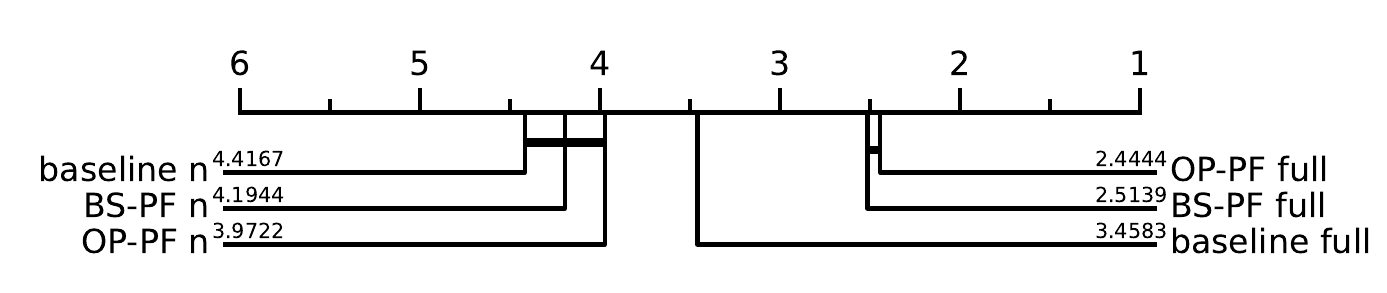}
        \caption{Simulated dataset}
        \label{fig:cd_plots:simulated}
    \end{subfigure}
    \caption{Critical difference plots showing the mean ranks of the approaches. `full' designates approaches using
    all elements. `n' designates those using only the mean motion.}
    \label{fig:cd_plots}
\end{figure}

\begin{figure}[ht!]
    \centering

    \begin{subfigure}[t]{0.49\linewidth}
        \centering
        \includegraphics[width = 0.85\linewidth]{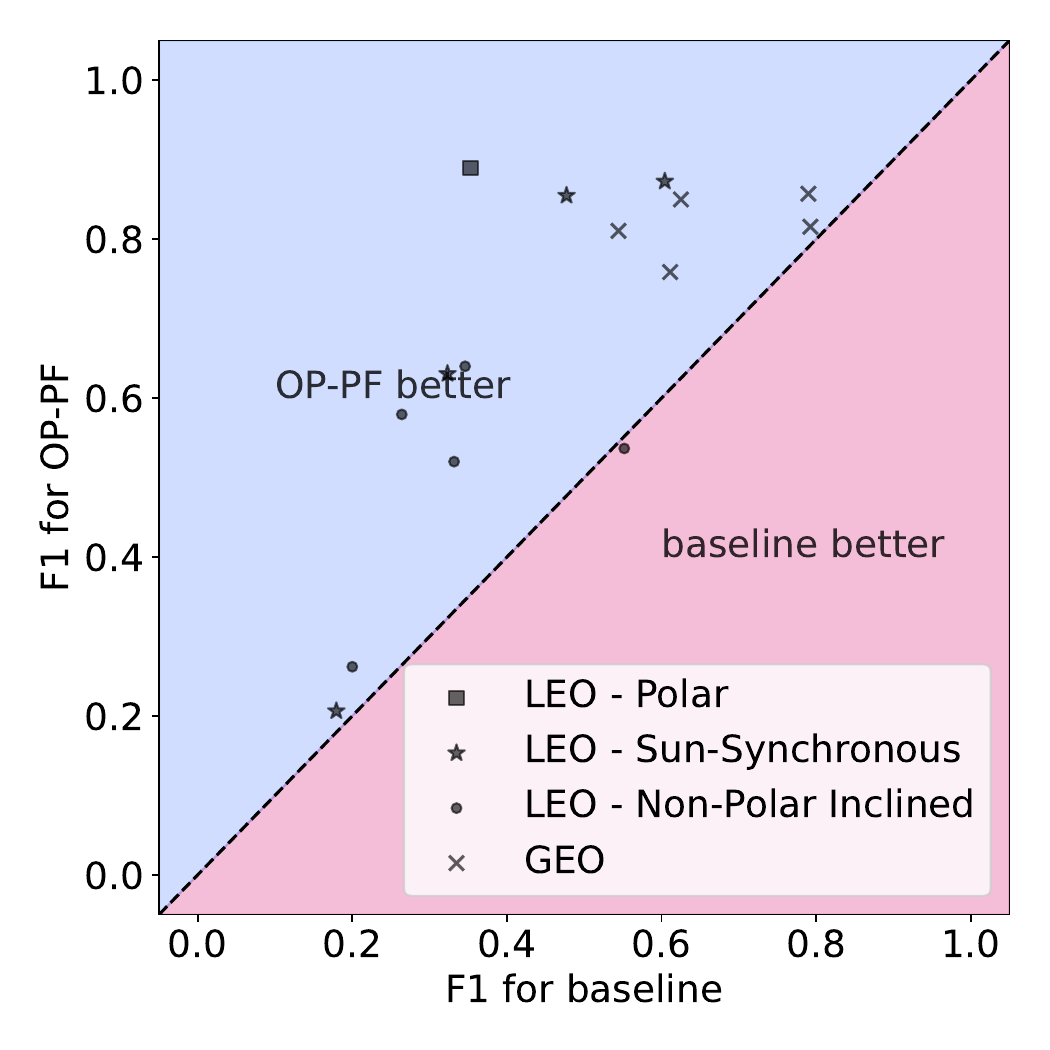}
        \caption{Benchmark dataset}
        \label{fig:results:benchmark}
    \end{subfigure} \hfil
    \begin{subfigure}[t]{0.49\linewidth}
        \centering
        \includegraphics[width = 0.85\linewidth]{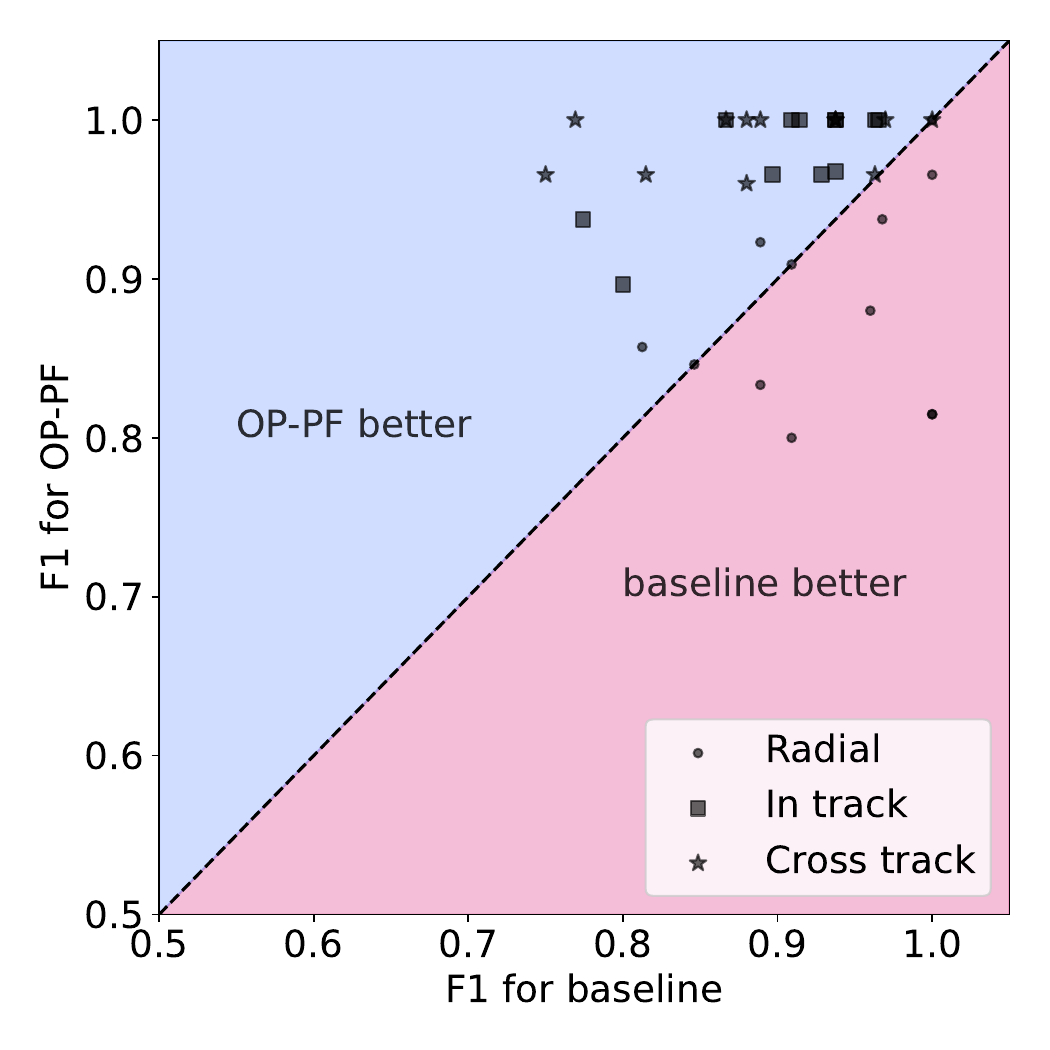}
        \caption{Simulated data}
        \label{fig:results:simulated}
    \end{subfigure} \hfil
    \caption{Comparative plots showing the F1 scores achieved by the baseline and OP-PF on the satellites in the
    benchmark and simulated test sets. All orbital elements were used by both approaches. The markers designate
    either the satellite's orbit type or the type of manoeuvres inserted into a simulation run.}
    \label{fig:results}
\end{figure}

We test the proposed methods on a recently-released benchmark dataset of satellite TLEs and independently-obtained
ground-truth manoeuvre timestamps~\cite{shorten2023wide, shorten2022wide}. This dataset comprises 15
satellites in a variety of orbits.

Manoeuvres in this dataset are associated with large changes in the mean motion. We exploit this by developing
versions of both the particle filters and the baseline which focus exclusively on this element. For the baseline
approach, the modified version using only the mean motion applies \eqRef{eq:baseline_anomaly_statistic} using only
this element. For the two particle filtering approaches, the marginal negative log of the predictive density
\eqRef{eq:predictive_density} is computed for the mean motion.

\figRef{fig:cd_plots:benchmark} plots the mean ranks of the six combinations of methods and inspected elements. The
ranks are determined by the F1 score. The thick horizontal lines indicate that there are no significant pairwise
differences in rank between any of the methods under the line according to a Wilcoxon-Holm test~\cite{ismail2019deep}.
When using all elements, the filtering approaches perform significantly better than the baseline. However, all
methods perform better when using only the mean motion and their performance in this case is similar.

\figRef{fig:results:benchmark} plots the performance of the Optimal-Proposal Particle Filter (OP-PF) against the
baseline, when using all elements, on each satellite. Each point in this plot represents a specific satellite and the
$x$ and $y$ values are the F1 score for the baseline and OP-PF, respectively. The OP-PF achieves a higher F1
score than the baseline on all satellites except one.
\vspace{-3mm}
\subsection{Evaluation on Simulated Data}
\label{sec:experiments:simulated}

To tease apart the relative performance of the approaches under investigation, we tested them on simulated
data with less obvious manoeuvres of varying types.
TLE orbit trajectories were simulated using SGP4 by starting at a TLE record of an existing
satellite (not in the benchmark set used in \secRef{sec:experiments:benchmark}) and then propagating to each
subsequent epoch time iteratively. Both propagation and observation noise were added in this process. At selected
time points, small impulsive manoeuvres in the radial, in-track or cross-track directions were added. All inserted
manoeuvres in each simulation run were of a single type. A total of 36 short (500 epochs) TLE
orbit trajectories were created in this manner.

In contrast to the results on the benchmark dataset presented in \secRef{sec:experiments:benchmark}, the strategy of
focussing on the mean motion alone is not effective on this simulated dataset. \figRef{fig:cd_plots:simulated} plots
the mean ranks of the six combinations of methods and inspected elements. The two filtering approaches using all
elements outperform all other approaches.

\figRef{fig:results:simulated} shows a comparative plot of the performance of the OP-PF and
the baseline when using all elements. The
OP-PF achieves superior performance on a majority of simulation runs. Moreover, the baseline only achieves superior
performance for simulation runs with only radial manoeuvres. \FloatBarrier

\vspace{-3mm}
\section{Discussion}
\label{sec:discussion}

We developed a particle filter approach to track the underlying orbital state of satellites from the noisy
measurements provided in TLE data, using SGP4 as the forward model. Special attention was paid to the structure of
the errors produced by SGP4 and associated uncertainty covariances for the model and observations were derived.
The proposed filtering approaches were evaluated on a benchmark dataset~\cite{shorten2023wide, shorten2022wide}
consisting of 15 satellites. The size of this evaluation set is substantially larger than in any previous work.
Moreover, it has the added advantage of containing independently-derived manoeuvre timestamps.

The performance of the filtering approaches was compared against a baseline algorithm which operates similarly
to a number of previously-proposed approaches~\cite{li2018new, mukundan2021simplified, zhao2014method}. It
was found that, for the benchmark dataset, both the baseline and filtering approaches benefited from performing
anomaly detection using only the mean motion. When using only this element, the performance
of all the techniques was similar, and each of them outperformed all approaches which used all elements.
However, when making anomaly predictions using all elements, both  techniques achieved superior performance
over the baseline.

The approaches were further evaluated on a challenging set of simulated data. Here, performing anomaly
detection using only the mean motion had a negative effect on all methods and the particle filters significantly
outperformed the baseline.

One limitation of the proposed approach is that it uses the given TLE set to estimate the covariance matrix of the
model and observation uncertainty prior to filtering. The filtering framework within which we are operating estimates
the distribution of the subsequent observation using the preceding observations
$p\left( \mathbf{y}_{k+1} \mid \mathbf{y}_{1:k}\right)$. However, our approach to deriving the model and observation
uncertainty uses the entire set of observations ($\mathbf{y}_{1:K}$), implying that some information from future
observations is used to estimate the predictive density. It is unlikely that this choice has much practical impact.
One alternate approach would be to use an initial portion of the data to estimate the covariance matrices.

A further limitation of the presented method for estimating the uncertainty is that it does not take into account any
time dependency of the model uncertainty. Although the published TLE epochs are usually spaced at roughly even
intervals~\cite{vallado2006revisiting}, this is not always the case. A straightforward approach to incorporating this
time dependency would be to group inter-epoch intervals into discrete bins and estimate a model uncertainty for each
bin.

\vspace{-3mm}
\section{Acknowledgement}

We thank Will Heyne of BAE Systems for helpful insights throughout the course of this research. This work has been
supported by the SmartSat CRC, whose activities are funded by the Australian Government’s CRC Program. 
\FloatBarrier

\bibliographystyle{splncs04}
\bibliography{PF_manoeuvre_detection}

\begin{thebibliography}{10}
\providecommand{\url}[1]{\texttt{#1}}
\providecommand{\urlprefix}{URL }
\providecommand{\doi}[1]{https://doi.org/#1}

\bibitem{celesTrak}
{CelesTrak}: {NORAD} two-line element set format.
  \url{celestrak.org/NORAD/documentation/tle-fmt.php}, accessed: 2022-10-24

\bibitem{arulampalam2002tutorial}
Arulampalam, M.S., Maskell, S., Gordon, N., Clapp, T.: A tutorial on particle
  filters for online nonlinear/non-{G}aussian {B}ayesian tracking. IEEE
  Transactions on signal processing  \textbf{50}(2),  174--188 (2002)

\bibitem{blanchard2021accurately}
Blanchard, P., Higham, D.J., Higham, N.J.: Accurately computing the log-sum-exp
  and softmax functions. IMA Journal of Numerical Analysis  \textbf{41}(4),
  2311--2330 (2021)

\bibitem{decoto2015technique}
Decoto, J., Loerch, P.: Technique for {GEO} {RSO} station keeping
  characterization and maneuver detection. In: Advanced Maui Optical and Space
  Surveillance Technologies Conference. p.~42 (2015)

\bibitem{doucet2000sequential}
Doucet, A., Godsill, S., Andrieu, C.: On sequential {Monte} {Carlo} sampling
  methods for {Bayesian} filtering. Statistics and Computing  \textbf{10}(3),
  197--208 (Jul 2000). \doi{10.1023/A:1008935410038}

\bibitem{geul2017tle}
Geul, J., Mooij, E., Noomen, R.: Tle uncertainty estimation using robust
  weighted differencing. Advances in Space Research  \textbf{59}(10),
  2522--2535 (2017)

\bibitem{hoots1980spacetrack}
Hoots, F.R., Roehrich, R.L.: Models for propagation of {NORAD} element set.
  Tech. Rep.~3, Aerospace Defense Command, United States Air Force (1980),
  spacetrack report No.3

\bibitem{ismail2019deep}
Ismail~Fawaz, H., Forestier, G., Weber, J., Idoumghar, L., Muller, P.A.: Deep
  learning for time series classification: a review. Data mining and knowledge
  discovery  \textbf{33}(4),  917--963 (2019)

\bibitem{kelecy2007satellite}
Kelecy, T., Hall, D., Hamada, K., Stocker, D.: Satellite maneuver detection
  using two-line element ({TLE}) data. In: Proceedings of the Advanced Maui
  Optical and Space Surveillance Technologies Conference. Maui Economic
  Development Board (MEDB) Maui, HA (2007)

\bibitem{kennewell2013overview}
Kennewell, J.A., Vo, B.N.: An overview of space situational awareness. In:
  Proceedings of the 16th International Conference on Information Fusion. pp.
  1029--1036. IEEE (2013)

\bibitem{leeuwen2019particle}
Leeuwen, P.J.v., Künsch, H.R., Nerger, L., Potthast, R., Reich, S.: Particle
  filters for high-dimensional geoscience applications: {A} review. Quarterly
  Journal of the Royal Meteorological Society  \textbf{145}(723),  2335--2365
  (2019). \doi{10.1002/qj.3551}

\bibitem{li2018new}
Li, T., Li, K., Chen, L.: New manoeuvre detection method based on historical
  orbital data for low earth orbit satellites. Advances in Space Research
  \textbf{62}(3),  554--567 (2018)

\bibitem{li2019maneuver}
Li, T., Li, K., Chen, L.: Maneuver detection method based on probability
  distribution fitting of the prediction error. Journal of Spacecraft and
  Rockets  \textbf{56}(4),  1114--1120 (2019)

\bibitem{mukundan2021simplified}
Mukundan, A., Wang, H.C.: Simplified approach to detect satellite maneuvers
  using {TLE} data and simplified perturbation model utilizing orbital element
  variation. Applied Sciences  \textbf{11}(21),  10181 (2021)

\bibitem{musso2001improving}
Musso, C., Oudjane, N., Le~Gland, F.: Improving regularised particle filters.
  In: Sequential Monte Carlo methods in practice, pp. 247--271. Springer (2001)

\bibitem{roberts2021geosynchronous}
Roberts, T.G., Linares, R.: Geosynchronous satellite maneuver classification
  via supervised machine learning. Tech. rep., Massachusetts Institute of
  Technology (2021)

\bibitem{san2017hybrid}
San-Juan, J.F., P{\'e}rez, I., San-Mart{\'\i}n, M., Vergara, E.P.: Hybrid sgp4
  orbit propagator. Acta Astronautica  \textbf{137},  254--260 (2017)

\bibitem{shorten2022wide}
Shorten, D.P., Yang, Y., Maclean, J., Roughan, M.: Wide-scale monitoring of
  satellite lifetimes: Pitfalls and a benchmark dataset. arXiv preprint
  arXiv:2212.08662  (2022)

\bibitem{shorten2023wide}
Shorten, D.P., Yang, Y., Maclean, J., Roughan, M.: Wide-scale monitoring of
  satellite lifetimes: Pitfalls and a benchmark dataset. Journal of Spacecraft
  and Rockets pp.~1--5 (2023)

\bibitem{snyder2011particle}
Snyder, C.: Particle ﬁlters, the “optimal” proposal and high-dimensional
  systems. ECMWF Seminar on Data assimilation for atmosphere and ocean p.~10
  (2011)

\bibitem{speekenbrink2016tutorial}
Speekenbrink, M.: A tutorial on particle filters. Journal of Mathematical
  Psychology  \textbf{73},  140--152 (2016)

\bibitem{ucs}
{Union of Concerned Scientists}: {UCS} satellite database.
  \url{www.ucsusa.org/resources/satellite-database}, accessed: 2023-08-21

\bibitem{vallado2006revisiting}
Vallado, D., Crawford, P., Hujsak, R., Kelso, T.: Revisiting spacetrack report
  \#3. In: AIAA/AAS Astrodynamics Specialist Conference and Exhibit. p.~6753
  (2006)

\bibitem{vallado2001fundamentals}
Vallado, D.A.: Fundamentals of astrodynamics and applications, vol.~12.
  Springer Science \& Business Media (2001)

\bibitem{vallado2012two}
Vallado, D.A., Cefola, P.J.: Two-line element sets - practice and use. In: 63rd
  International Astronautical Congress, Naples, Italy. pp. 1--14 (2012)

\bibitem{wang2021gaussian}
Wang, Y., Bai, X., Peng, H., Chen, G., Shen, D., Blasch, E., Sheaff, C.B.:
  Gaussian-binary classification for resident space object maneuver detection.
  Acta Astronautica  \textbf{187},  438--446 (2021)

\bibitem{zhao2021event}
Zhao, L.: Event prediction in the big data era: A systematic survey. ACM
  Computing Surveys (CSUR)  \textbf{54}(5),  1--37 (2021)

\bibitem{zhao2022overview}
Zhao, Q., Yu, L., Du, Z., Peng, D., Hao, P., Zhang, Y., Gong, P.: An overview
  of the applications of earth observation satellite data: impacts and future
  trends. Remote Sensing  \textbf{14}(8), ~1863 (2022)

\bibitem{zhao2014method}
Zhao, Y., Zhang, K., Bennett, J., Sang, J., Wu, S.: A method for improving
  two-line element outlier detection based on a consistency check. In:
  Proceedings of the Advanced Maui Optical and Space Surveilllance (AMOS)
  Technologies Conference, Maui, Hawaii (2014)

\end{thebibliography}

\end{document}